\documentclass[baaa]{baaa}
\usepackage[pdftex]{hyperref}
\usepackage{subfigure}
\usepackage{natbib}
\usepackage{helvet,soul}
\usepackage[font=small]{caption}
\contriblanguage{1}

\contribtype{1}

\thematicarea{4}

\received{\ldots}
\accepted{\ldots}

\title{Bayesian characterization of the young open cluster NGC 6383 using HDBSCAN and Gaia DR3}

\titlerunning{Bayesian characterization of NGC 6383}

\author{
L.M. Pulgar-Escobar \inst{1} \& N.A. Henríquez-Salgado\inst{1}}

\authorrunning{Pulgar-Escobar et al.}

\contact{lescobar2019@udec.cl}

\institute{Departamento de Astronom{\'\i}a, Universidad de Concepci\'on, Chile
}

\resumen{Este estudio se centra en determinar las características del joven cúmulo abierto NGC 6383. Para lograrlo, se utiliza el algoritmo de agrupamiento \sc HDBSCAN\rm\,para identificar los miembros potenciales del cúmulo basándose en los movimientos propios y los paralajes utilizando \sl Gaia Data Release 3\rm. Se obtienen varios parámetros de NGC 6383, como el \textit{tidal radius}, el radio del núcleo, la distancia mediante paralaje y ajuste de isocronas, el movimiento propio promedio, la edad, la metalicidad y otros relevantes. Para realizar este análisis, utilizamos una extensión de Monte Carlo Hamiltoniano, el \textit{No-U-Turn Sampler}. Los resultados de este análisis señalan que NGC 6383 es un cúmulo abierto muy joven $(\approx 1 - 4~\mathrm{Myr})$, con una distancia de $\approx 1.1~\mathrm{kpc}$.}

\abstract{This study focuses on determining the characteristics of the young open cluster NGC 6383. To achieve this, the \sc HDBSCAN\rm\,clustering algorithm is utilized to identify potential cluster members based on proper motions and parallaxes from \sl Gaia Data Release 3\rm. Various parameters of NGC 6383, such as tidal radius, core radius, distance through parallax and isochrone-fitting, proper motion, age, metallicity, and relevant others, are assessed. To perform this analysis, we utilize an extension of Hamiltonian Monte Carlo, the No-U-Turn Sampler. The results of this analysis point out that NGC 6383 is a very young open cluster $(\approx 1 - 4~\mathrm{Myr})$, with a distance of $\approx 1.1~\mathrm{kpc}$.}


\begin{document}
\maketitle
\section{Introduction}\label{S_intro}
Precise cluster’s members identification is a crucial step when it comes to a correct identification and characterization of the cluster. Various methods have been used for this process  \citep{ 1968ArA.....5....1L,1978MNRAS.182..607F, 1989MNRAS.236..263P, 2005AA...438.1163K, 2007AA...462..157P}, each offering different focuses and sometimes leading to consensus, or discrepancies. Recently, Bayesian analysis implementation on the membership identification, offers a different method to compare or compliment with the ones mentioned.

NGC 6383 \footnote{NGC 6383, also known as NGC 6374 in the New General Catalog and classified as Collinder 334 and Collinder 335 in the Collinder catalog, was initially misclassified in the original Collinder catalog \citep{1931AnLun...2....1C}.} is a young open cluster situated in the Carina-Sagittarius arm, within the Sh 2-012 star formation region. It forms part of the larger Sagittarius OB1 association, along with NGC 6530 and NGC 6531. The galactic coordinates of the cluster are $\ell = 355.68^{\circ}$ and $b = 0.05^{\circ}$ \citep{2008hsf2.book..497R}. 

The distance to the cluster has been estimated by various authors, with an upper limit of $2130 ~\mathrm{pc}$ proposed by \cite{1930LicOB..14..154T} and \cite{1937LicOB..18...89Z}, and a lower limit of $760~\mathrm{pc}$ and $834~\mathrm{pc}$ suggested by \cite{1949ApJ...110..117S} and \cite{2018AA...610A..30Arib}, respectively.

The parameters of the cluster exhibit a wide range of results, contingent on the methodology utilized by the authors in their research. Consequently, the age of the cluster has been a topic of considerable debate.

The implementation of Bayesian analysis and Machine Learning techniques in this study, explores a different method to obtained various parameters for characterizing NGC 6383 and contribute to the debate.
\section{Methodology}
We acquired data from the \textit{Gaia} third Data Release (DR3), executing a cone search of $25~\mathrm{arcmin}$ radius on the \textit{Gaia} archive, yielding $71847$ sources. Initial selection filtered sources to distances between $500~\mathrm{pc}$ and $2500~\mathrm{pc}$—equivalent to parallax ranges between $0.4~\mathrm{mas}$ and $2~\mathrm{mas}$—based on limits established in prior research, resulting in $29092$ sources.

\begin{figure}
    \centering
    \includegraphics[width=\columnwidth]{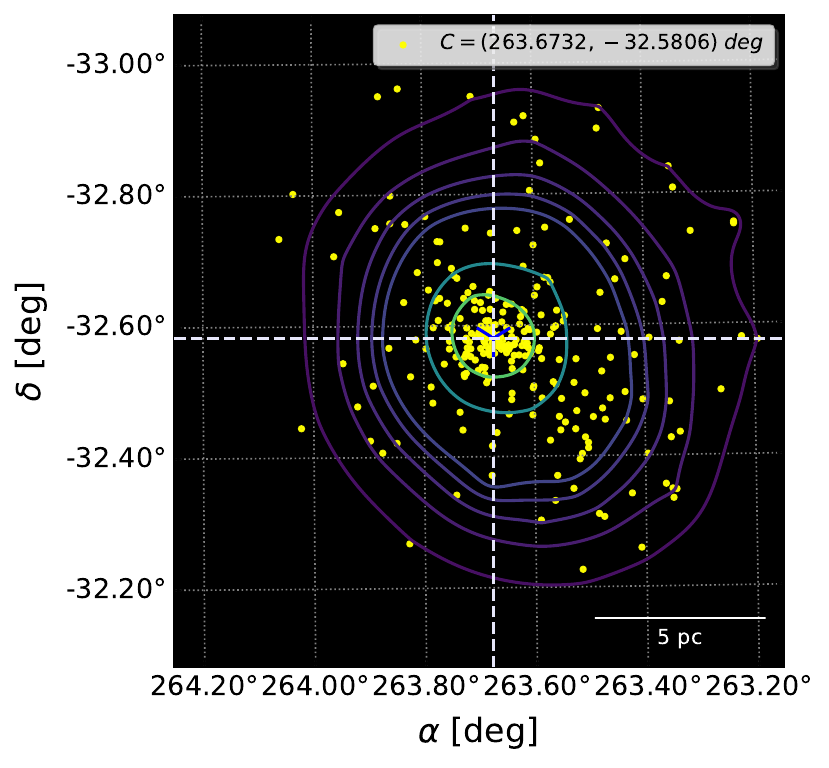}
    \caption{Spatial distribution of sources in R.A. ($\alpha$) and DEC. ($\delta$), considering a probability of at least $60$ percent. Concentric contour lines indicate levels of the KDE with exponential kernel, and the center of the distribution is marked by the white dashed lines.}
    \label{fig:center}
\end{figure}
To enhance data reliability, we applied astrometric fidelity parameters from \cite{2022MNRAS.510.2597R}. This parameter, derived from a neural network analysis of $17$ \textit{Gaia} catalog metrics, assesses the astrometric solution's trustworthiness. Sources with an astrometric fidelity above $0.5$ were retained, narrowing the selection to $20215$ sources.

Subsequent refinement ensured inclusion of only sources with comprehensive parameters, finalizing the dataset at $19964$ sources. Systematic parallax offsets identified in \citet{2021A&A...649A...2L} were corrected using the \textsc{Gaiadr3\_zeropoint} package. Moreover, addressing the bias in proper motion for bright sources ($\mathrm{G}<13~\mathrm{mag}$), as discussed by \citet{2021A&A...649A.124C}, we applied a magnitude-based correction for sources with $\mathrm{G} = 11-13~\mathrm{mag}$, compensating for up to $80~\mathrm{\mu as\,yr^{-1}}$ discrepancy between the frames of reference for bright and faint sources.
\subsection{\sc COSMIC}
Characterization Of Star clusters using Machine learning Inference and Clustering (\textsc{COSMIC}) developed by Lucas Pulgar-Escobar et al. (in prep), is a suite of functions designed for analyzing open clusters. Utilizing unsupervised machine learning algorithms, \textsc{COSMIC} processes extensive datasets, such as those from \textit{Gaia}, to identify fundamental parameters of open clusters through clustering techniques and Bayesian estimation. As an open-source program, it is developed in \textsc{Python} 3.11 and integrates \textsc{PyMC} 5.10\footnote{\href{www.pymc.io}{www.pymc.io}}, a \textsc{Python} library specialized in Bayesian analysis.
\begin{figure}
    \centering
    \includegraphics[width=\columnwidth]{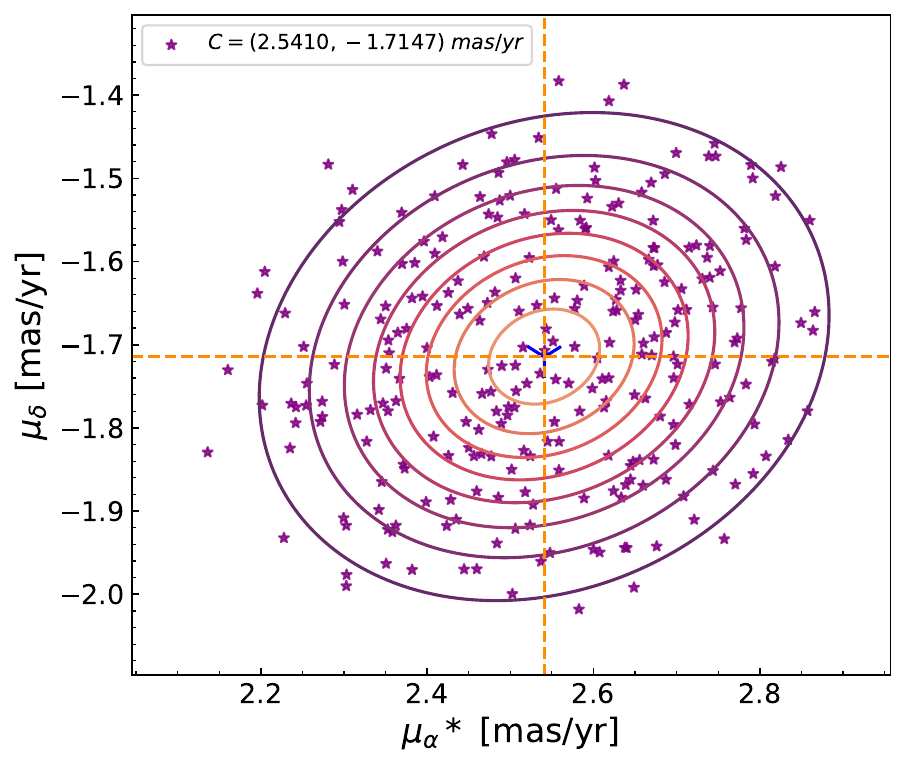}
    \caption{Proper motions of sources within the cluster in R.A. ($\mu_\alpha^*$) and DEC. ($\mu_\delta$), considering a probability of at least $60$ percent. Concentric contour lines indicate levels of the multivariate Gaussian, and the center of the proper motion distribution is marked by the orange dashed lines.}
    \label{fig:proper_motion}
\end{figure}
\subsubsection{Membership determination}
To identify potential members of NGC 6383, we employed the Hierarchical Density-Based Spatial Clustering of Applications with Noise (\textsc{HDBSCAN}) algorithm \citep{10.1007/978-3-642-37456-2_14}, known for its effectiveness in identifying clusters of varying densities and sizes without requiring a predetermined number of clusters \citep{2021A&A...646A.104H}. This density-based clustering method, which constructs a hierarchical representation of the data \citep{McInnes2017}, is particularly adept at handling different cluster shapes and sizes while distinguishing noise points. Its capacity to automatically determine the optimal cluster count makes \textsc{HDBSCAN} an ideal choice for this analysis, offering significant benefits over traditional clustering algorithms, especially in terms of robustness against noisy data and outliers.
\begin{figure*}
    \centering
    \includegraphics[width=\textwidth]{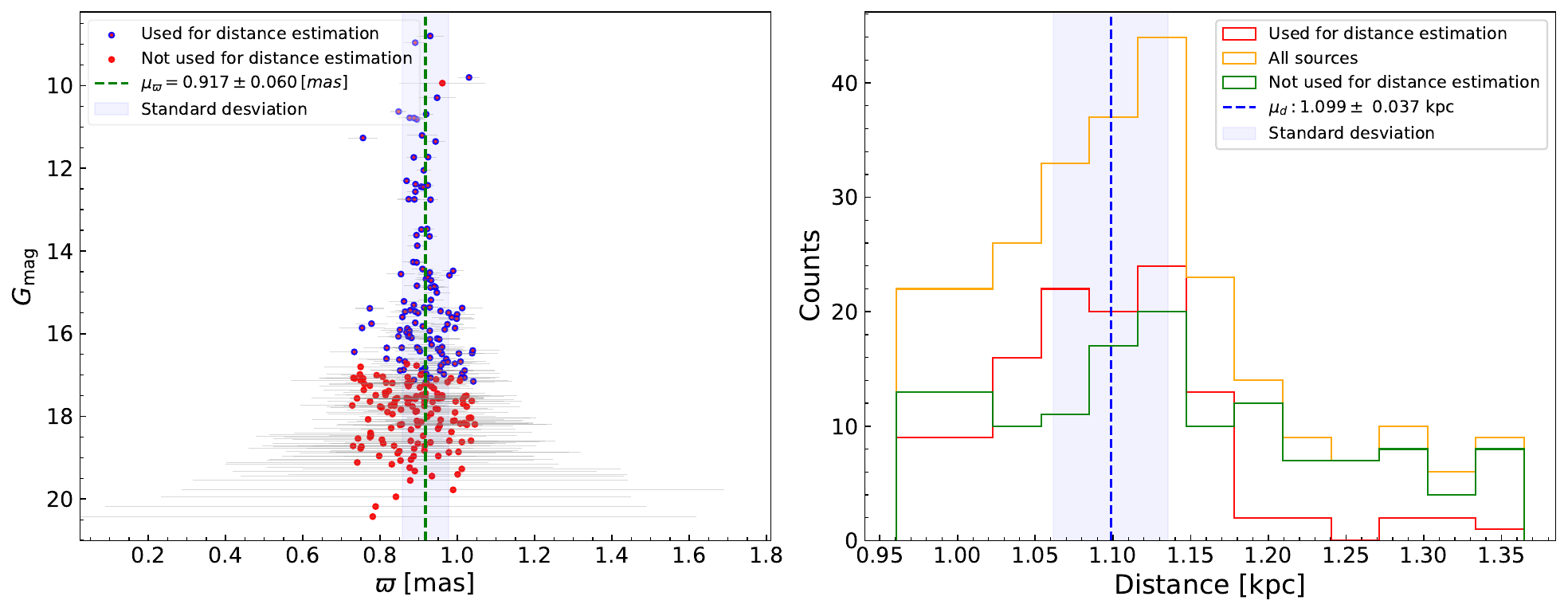}
    \caption{\emph{Left panel:} \sl Gaia \rm G-band magnitude ($G_{\mathrm{mag}}$) vs. parallax ($\varpi$), it's noticeable that fainter sources exhibit larger parallax errors, rendering them less reliable. Blue dots are sources with fractional parallax less than $0.1$, and red ones are sources with values over $0.1$, being excluded for the distance estimation. The green dashed line indicates mean parallax.
    \emph{Right panel:} Histogram of the inverse values of the parallaxes as distances, with the same criteria as the left panel.}
    \label{fig:parallax_distance}
\end{figure*}
\textsc{HDBSCAN} was applied\footnote{The algorithm's hyperparameters included an Euclidean distance metric and a minimum cluster size of $43$.} to the $19964$ sources using the proper motions as the clustering parameters, resulting in 544 sources with a probability over $0.5$.

We applied the \textsc{AstroPy Sigma Clipping} utility \citep{astropy:2022} for a $2\sigma$ clipping around the median, yielding 399 probable members for parameters inference.
\subsubsection{Parallax and Distance Estimation}
Distance estimation from parallax is challenging due to measurement errors, and transforming parallax to distance is not straightforward and involves significant uncertainties, deviating from the simple \textit{inverse of the parallax} approach \citep{2015PASP..127..994B,2021AJ....161..147B}. A crucial aspect is choosing an appropriate prior distribution. When analyzing the parallax measurements of stars within an open cluster and assuming a simplified one-dimensional perspective, we can see that the distribution of this parallax values follows a normal distribution. For accuracy, our analysis only includes members with fractional parallax errors under $0.1$.

Our hierarchical model begins by calculating an initial average distance $(\mu_{\mathrm{prior}})$ from the observed parallaxes. The cluster's mean distance follows $\mathcal{U}(0.5\mu_{\mathrm{prior}},1.5\mu_{\mathrm{prior}})$ as a prior.
\subsubsection{Proper motions}
To estimate the distribution of cluster members' proper motions, we model them using a two-dimensional normal distribution. 

A normal prior distribution was assigned to the mean proper motions, based on frequentist means and standard deviations. The standard deviations are modeled using a HalfNormal distribution, while the correlation coefficient between the proper motions follows a $\mathcal{U}(-1, 1)$.
\begin{figure}
    \centering
    \includegraphics[width=\columnwidth]{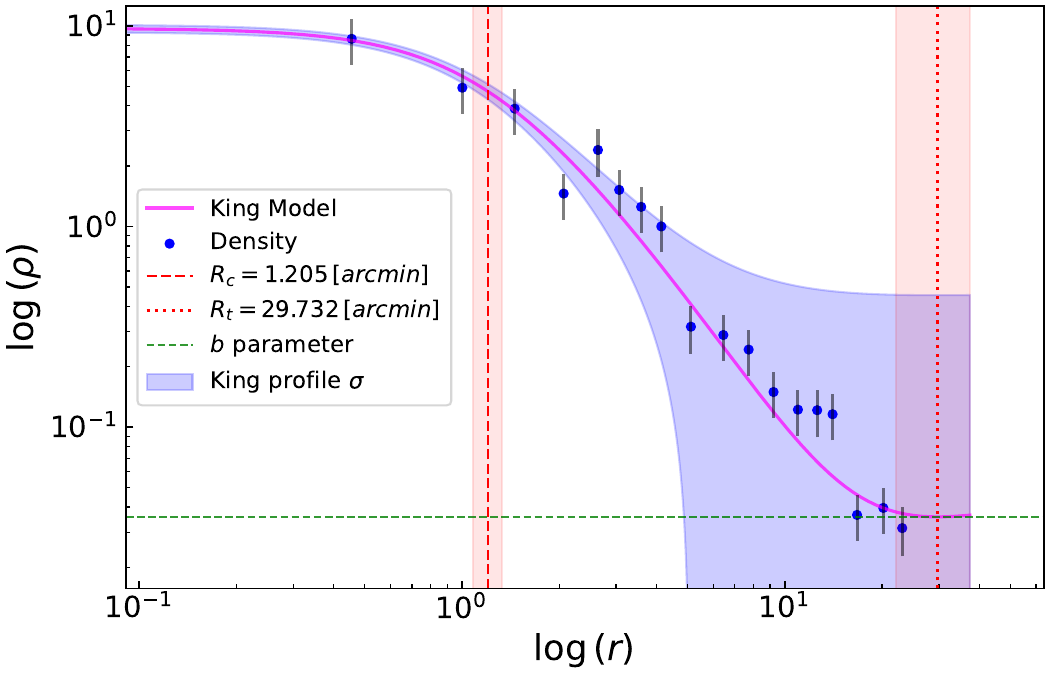}
    \caption{Radial density profile of NGC 6383, with the blue dots representing the stellar density. The shaded blue zone outlines the deviation of the King's profiles, and the magenta line represents the median profile. The core radius $R_c$ is shown as a red dashed line, the tidal radius $R_t$ as a red dotted line, and the background density $b$ as a green dashed line. The logarithmic axes denote the radius $r$ in arcminutes and the stellar density $\rho$ per arcminute squared.}
    \label{fig:king_profile}
\end{figure}
\subsubsection{Center determination}
To  determine the cluster's center, we used a weighted Kernel Density Estimation (KDE) from \textsc{sci-kit learn} \citep{scikit-learn}, assigning weights inversely proportional to the distance from the mean proper motion. Optimal KDE parameters were determined through Grid Search Cross-Validation within the package, exploring a range of bandwidths—from the mean positional error to the search cone radius—and all kernel types. The location of maximum density was designated as the cluster's center.
\subsubsection{Radial density profile}
To determine structural parameters of the cluster, we utilized the \citet{1962AJ.....67..471K} density profile. The numerical density $\rho$, was computed by dividing the cluster area into concentric annuli, each containing an equal number of stars. The number of annuli, $K$, adheres to the equiprobable bin rule ($K = 2n^{2/5}$), where $n$ is the star count within the cluster.

Model priors were set as follows: background density $b \sim \mathcal{U}(0, 2\rho_{\mathrm{min}})$, scale factor $k \sim \mathcal{U}(0, 2\rho_{\mathrm{max}})$, core radius $R_c \sim \mathcal{U}(0, 0.8R_t)$, and tidal radius $R_t \sim \mathcal{U}(R_c, 1.5T_{\mathrm{max}})$. $T_{\mathrm{max}}$ is the maximum value among tidal potential radius, Hill radius, gravitational bound radius, and the maximum observed cluster member distance. This ensures the chosen parameter space for $R_c$ and $R_t$ is within astrophysical valid limits.

\subsection{AsTeCA}
AsTeCA, or Automated Stellar Cluster Analysis \citep{2015A&A...576A...6P}, is a suite of tools designed to automate the standard tests applied to stellar clusters to determine their basic parameters. In order to obtain accurate estimates for a cluster's metallicity, age and extinction values, we use AsTeCA's isochrone fitting process. We used the \textsc{PARSEC v1.2S} isochrones \citep{2012MNRAS.427..127B,2014MNRAS.445.4287T}, with the \textsc{Gaia} EDR3 photometric system and \citet{2001MNRAS.322..231K,2002Sci...295...82K} canonical two-part-power law IMF corrected for unresolved binaries, with a logarithmic age range between $6.0$ and $7.9$. All parameter priors are set as uniform distributions, and the median values of the resulting posterior distributions are informed.
\begin{table}[!t]
\centering
\caption{Obtained results for the cluster, distance($\varpi$) is the distance obtained by parallax analysis. On the other hand, D.M. indicates distance distance modulus.}
\begin{tabular}{llc}
\hline\hline\noalign{\smallskip}
\textbf{Parameter} & \textbf{Value} & \textbf{Unit} \\ \hline\noalign{\smallskip}
Distance $(\varpi)$ & $1.099 \pm 0.037$ & $\mathrm{kpc}$ \\
Distance (D.M.) & $1.534 \pm 0.258$ & $\mathrm{kpc}$ \\
Age & $6.202 \pm 0.036$ & $\log(\mathrm{age})$ \\
Metallicity ($Z$) & $0.015 \pm 0.014$ & - \\
Parallax $(\varpi)$ & $0.917 \pm 0.060$ & $\mathrm{mas}$ \\
Number of members & $266$ & $\mathrm{stars}$ \\
Absorption $(A_V)$ & $1.468 \pm 0.081$ & $\mathrm{mag}$ \\
Core Radius $(R_c)$ & $1.205 \pm 0.126$ & $\mathrm{arcmin}$ \\
Background $(b)$ & $0.035 \pm 0.017$ & $\mathrm{stars\,arcmin^{-2}}$ \\
Distance Modulus & $10.903 \pm 0.327$ & $\mathrm{mag}$ \\
Tidal Radius $(R_t)$ & $29.732 \pm 7.694$ & $\mathrm{arcmin}$ \\
Center Density $(k)$ & $10.508 \pm 0.795$ & $\mathrm{stars\,arcmin^{-2}}$ \\
Cluster Center R.A. & $263.673 \pm 0.004$ & $\mathrm{deg}$ \\
Cluster Center DEC. & $-32.580 \pm 0.004$ & $\mathrm{deg}$ \\
Proper Motion R.A. & $2.541 \pm 0.007$ & $\mathrm{mas\,yr^{-1}}$ \\
Proper Motion DEC. & $-1.714 \pm 0.006$ & $\mathrm{mas\,yr^{-1}}$ \\ \hline
\end{tabular}
\label{tab:ngc6383_results}
\end{table}
\begin{figure}
    \centering
    \includegraphics[width=\columnwidth]{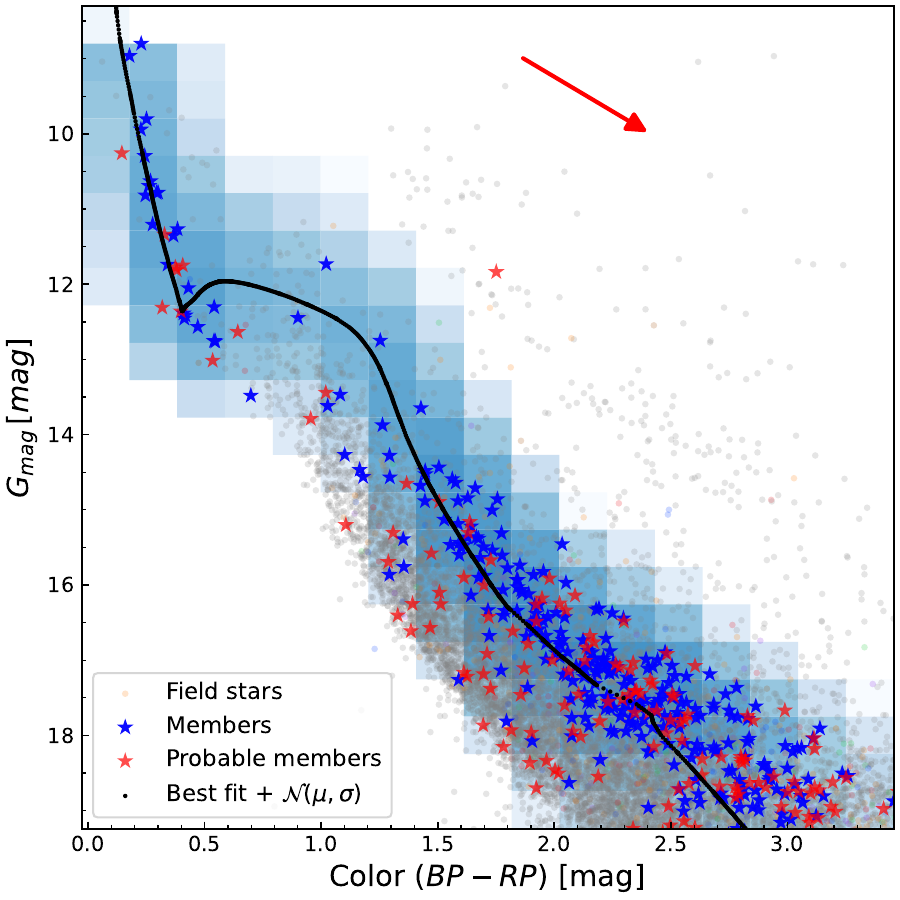}
    \caption{Color-Magnitude Diagram of the cluster. The best isochrone fitting obtained by AsTeCa is indicated with a black line. Blue stars are sources with probabilities over 60 percent, while red stars are sources with probabilities under 60 percent. The red arrow is the reddening vector. The shaded area is a normal distribution around the best fit.}
    \label{fig:cmd}
\end{figure}
\section{Preliminary Results}
The informed results and figures are obtained using sources with a membership probability exceeding 60 percent, ensuring a high level of confidence in the analysis. Table \ref{tab:ngc6383_results} presents a summary of the results, highlighting key parameters of NGC 6383. The cluster's center is accurately defined in Fig.\ref{fig:center}, while Fig.\ref{fig:proper_motion} illustrates the proper motions. Insights into distance and parallax estimation are provided in Fig.\ref{fig:parallax_distance}, and Fig.\ref{fig:king_profile} showcases the cluster's structural parameters alongside the fitted King's Profile. Additionally, Fig.\ref{fig:cmd} displays the CMD of NGC 6383 alongside the best-fit isochrone.

\section{Conclusion and future work}
Our study employed Bayesian methods and \textsc{HDBSCAN} to refine parameters like distance, proper motion, and age for NGC 6383 using Gaia DR3. 

Future efforts will broaden member analysis for a complete census and incorporate radial velocity for dynamic insights. Also, the use of a larger search cone to find possible stars in the tidal region of the cluster.  Additionally, incorporating multi-wavelength data could enhance the accuracy of determining stellar properties. Finally, a focused investigation into the binary and multiple systems within NGC 6383 like the central binary HD 159176 could offer valuable information about its star formation history.
\begin{acknowledgement}
We gratefully acknowledge support by the ANID BASAL project FB210003.
\end{acknowledgement}


\bibliographystyle{baaa}
\small
\bibliography{bibliografia}
 
\end{document}